\documentclass[12pt,aps,prl,preprint,tightenlines,
   showpacs,nofootinbib]{revtex4-1}
\newcommand{\PRE}[1]{{#1}} 

\usepackage{bm} \usepackage{epsfig}
\usepackage{slashed}
\usepackage{multirow}
\usepackage{amsmath}

\newcommand{\bea}{\begin{eqnarray}}
\newcommand{\eea}{\end{eqnarray}}

\begin{document}

\preprint{UCI-TR-2010-27}

\title{
\PRE{\vspace*{1.5in}} Anomaly breaking of de Sitter symmetry
\PRE{\vspace*{0.3in}} }

\author{Myron Bander}
\affiliation{Department of Physics and Astronomy, University of
California, Irvine, CA 92697, USA \PRE{\vspace*{.5in}} }

\date{November,\ \   2010}

\begin{abstract}
\PRE{\vspace*{.3in}} 
To one loop order, interacting boson fields on de Sitter space have an ``infrared" anomaly that breaks the de Sitter symmetry for all vacua 
save the Euclidian one.  The divergence of a symmetry current at point $x$ has a non-zero contribution at the antipodal point ${\bar x}$.
\end{abstract}


\maketitle

Quantum field theory on a de Sitter background presents problems primarily in the choice of vacuum on which to build the theory \cite{Mottola:1984ar,Allen:1985ux}.
Among these vacua the Euclidian or Bunch-Davies vacuum \cite{Bunch:1978yq} seems to presents least difficulties \cite{Einhorn-Larsen, Chernikov}. Recently Polyakov \cite{Polyakov:2007mm, Polyakov:2009nq} has argued that the criterion for which vacuum to use should be based on the behavior of the propagator at large distances.  Namely the propagator for a particle of mass $m$ should vary as  $\exp(-iml)$ , $l$ being the geodesic distance between two points in de Sitter space, rather than a sum of $\exp(iml)$ and $\exp(-iml)$; the latter is the behavior in all vacua save the one advocated in \cite{Polyakov:2007mm, Polyakov:2009nq}. The consequences of using this propagator are quite dramatic. The vacuum radites particles in an explosive way canceling the curvature of the underlying space. By screaning this curvature de Sitter symmetry is broken. This would have severe consequences for our understanding of cosmology, especially during inflationary growth. This picture was confirmed in the case of a two dimensional, (1+1), space \cite{Bander:2010pn} where by the use of a fermion-boson correspondence certain interacting theories can be solved exactly. It was found that a massless field with a sine-Gordon interaction correspond to a free fermion one with a de Sitter time dependent mass, explicitely breaking de Sitter symmetry. In this article we study the symmetries of interacting scalar theories in $D=(1+d)$ dimensional de Sitter space directly.  We find that, to one loop, there is an anomaly and the currents of de Sitter symmetry are {\em not} conserved. 

A D-dimensional de Sitter space, with coordinates $\tau, x_1,\cdots, x_d$ ($d=D-1)$) may be imbedded in a flat (D+1) Minkowski space with coordinates $Y_0, Y_1,\cdots, Y_D$, satisfying the constraint
\begin{equation}\label{imbed}
Y_0^2-Y_1^2-\cdots -Y_D^2=-1\,  .
\end{equation}
The parametrization we shall use is the flat slicing one \cite{Spradlin:2001pw} with conformal time where the metric of the space is
\begin{equation}\label{metric}
ds^2=d\tau^2-e^{2\tau}d{\vec x}d{\vec x}\, .
\end{equation}
The relation between the intrinsic de Sitter coordinates $\tau, {\vec x}$ and the embedding ones $Y_0,{\vec Y},Y_D$ (${\vec Y}$ denotes a $d$ dimensional vector) are
\begin{eqnarray}\label{conformalparametrization}
Y_0&=&\frac{1}{2}\left(\tau-\frac{1}{\tau}-\frac{{\vec x}^2}{\tau}\right)\, ,\nonumber\\
Y_i&=&-\frac{x_i}{\tau}\ \ \ (i=1\cdots d)\, ,\\
Y_D&=&\frac{1}{2}\left(-\tau-\frac{1}{\tau}+\frac{{\vec x}^2}{\tau}\right)\, .\nonumber
\end{eqnarray}
We are interested in how the isometries of de Sitter space are implemented in this metric. In the embedding space these isometries are Lorentz transformations involving $Y_0,{\vec Y}$ and $Y_D$ and fall into four classes: (i) velocity transformations in the $Y_i$ directions, (ii) velocity transformation in the $Y_D$ direction, (iii) rotations in the $Y_i-Y_j$ planes, and (iv) rotations in the $Y_i-Y_D $ planes; the infinitesimal forms of these and the coreponding transformations for the conformal $\tau,{\vec x}$ coordinates are:
\begin{widetext}
\begin{subequations}
\begin{eqnarray}
\delta Y_0=\epsilon Y_i;\ \ \delta Y_i=\epsilon Y_0 &\Longrightarrow & \delta\tau=-\tau x_i:\ \ \delta x_i= -\epsilon [x_ix_j+
\delta_{ij}(\tau^2-1-{\vec x}\cdot{\vec x})/2]\, , \label{dSsa}\\
\delta Y_0=\epsilon Y_D;\ \ \delta Y_D=\epsilon Y_0 &\Longrightarrow & \delta\tau=-\epsilon\tau;\ \ \delta X_i=-\epsilon x_i \label{dSsb}\, ,\\
\delta Y_i=\epsilon Y_j;\ \ \delta Y_j=-\epsilon Y_i; &\Longrightarrow &\delta x_i=\epsilon x_j;\ \ \delta x_j=-\epsilon x_i\label{dSsc}\, ,\\
\delta Y_D=\epsilon Y_i;\ \ \delta Y_i=-\epsilon Y_D &\Longrightarrow & 
\delta\tau=-\tau x_i;\ \ \delta x_j=-\epsilon [x_ix_j+\delta_{ij}(\tau^2+1-{\vec x}\cdot{\vec x})/2]\, ;\label{dSsd}\, .
\end{eqnarray}
\end{subequations}
\end{widetext}
An interacting scalar field, $\phi(\tau,{\vec x})$  propagating on a space with the metric in eq.~(\ref{metric}) is governed by the action
\begin{equation}\label{action}
{\cal S}=\int d\tau d^dx \tau^{-D}\left[\frac{\tau^2}{2}\partial^\mu\phi\phi\partial_\nu\phi-V(\phi)\right]\, ;
\end{equation}
indices are raised and lowered by a  $D$ dimensional flat Minkowski metric tensor $\eta^{\mu\nu}$. Details of calculations will be presented for the case where 
\begin{equation}\label{potential}
V(\phi)=\frac{m^2}{2}\phi^2+\frac{g}{4!}\phi^4\, 
\end{equation}
and then generalized to arbitrary $V(\phi)$.

The generators of the de Sitter isometries,  (\ref{dSsa}--\ref{dSsd}), are expressible in terms of the canonical energy-momentum tensor obtained from the Lagrangian in eq.~(\ref{action})
\begin{equation}\label{en-mom-tensor}
\Theta_{\mu\nu}=\tau^{2-D}\partial_\mu\phi\partial_\nu\phi-\eta_{\mu\nu}\left[\tau^{2-D}\frac{1}{2}
\partial_\alpha\phi\partial^\alpha\phi-\tau^{-D}V(\phi)\right]\, ;
\end{equation}
and are $S_\nu=\delta\tau \Theta_{0\nu}-\sum_i\delta x_i \Theta_{i\nu}$, or more specifically
\begin{widetext}
\begin{subequations}
\begin{eqnarray}
S^{(a;i)}_\nu&=&-\tau x_i\Theta_{0\nu}+\sum_j [x_ix_j+\delta_{ij}(\tau^2-1-{\vec x}\cdot{\vec x})/2]\Theta_{j\nu}\, ,\label{dSgena}\\
S^{(b)}_\nu&=&-\tau\theta_{0\nu}+\sum_jx_j\Theta_{j\nu}\, ,\label{dSgenb}\\
S^{(c;i,j)}_\nu&=&x_i\Theta_{j\nu}-x_i\Theta_{i\nu}\, ,\label{dSgenc}\\
S^{(d;i)}_\nu&=&-\tau x_i\Theta_{0\nu}+[x_ix_j+\delta_{ij}(\tau^2+1-{\vec x}\cdot{\vec x})/2]\Theta_{j\nu}\, .\label{dSgend}
\end{eqnarray}
\end{subequations}
\end{widetext}
The explicit appearance of the coordinate $\tau$ in  $\Theta_{\mu\nu}$ results in nonconservation of this tensor, however the condition
\begin{equation}\label{conscond}
\tau\partial^{\alpha}\Theta_{0\alpha}=\Theta^{\alpha}_{\ \alpha}
\end{equation}
ensures the conservation of all the de Sitter currents (\ref{dSgena}--\ref{dSgend}), i.e. $\eta^{\mu\nu}\partial_\mu S^{(..)}_\nu=0$

Using the equations of motion obtained from (\ref{action}), it is strairghtforward to check that the energy-momentum tensor, eq.~(\ref {en-mom-tensor}), satisfies (\ref{conscond}).  We shall show that for the propagator advocated in \cite{Polyakov:2007mm,Polyakov:2009nq} this relation does {\em not} hold for regulated one loop corrections. Explicitely, the propagator we shall use is 
\begin{equation}\label{polyakovprop}
D_m(x_1,x_2)=C(1-z_{12}^2)^{-(D-2)/4}{\cal Q}_{-\frac{1}{2}+i\nu(m)}^{(D-2)/2}(z_{12})\, ;
\end{equation}
with $Y_i$ the coordinates in the embedding space corresponding to the point $x_i$ in the de Sitter space; $z_{ij}=Y_i\cdot Y_j$. $\cal Q$ is an associated Legendre function of the second kind and $\nu(m)=\sqrt{m^2-(D-1)^2/4}$ and the constant $C$ depends only on the dimension of the de Sitter space and is chosen to insure a correct residue at $z_{12}=1$, corresponding to $Y_1=Y_2$. In addition to the ``ultraviolet" singularity at $Y_1=Y_2$ (\ref{polyakovprop}) has an additional, ``infrared", singularity at $z_{12}=-1$, namely $Y_2=-Y_1$ or $x_2={\bar x}_1$, the point antipodal to $x_1$ \cite{Spradlin:2001pw}.  It is this singularity that will be responsible for the non-conservation of de Sitter currents.

To determine the conservation, or lack thereof,  we shall study the matrix element
\begin{equation}\label{matelem}
T_{\mu\nu}(x; y_1, y_2)=\langle T[\Theta_{\mu\nu}(x)\phi(y_1)\phi(y_2)]\rangle\, ,
\end{equation}
where the symbol $T$ in the matrix element above indicates a conformal time, $\tau$, ordered product. To zoroth order in $g$ we find
\begin{widetext}
\begin{eqnarray}\label{0-order}
T_{\mu\nu}(x; y_1, y_2)&=&\left[\tau^{2-D}\partial_\mu D_m(x,y_1)\partial_\nu D_m(x,y_2)+(y_1\leftrightarrow y_2)\right]\nonumber\\
&-&\eta_{\mu\nu}\left[\tau^{2-D}\partial^\alpha D_m(x,y_1)\partial_\alpha D_m(x,y_2)-\tau^{-D}m^2D_m(x,y_1)D_m(x,y_2)\right]\, ,
\end{eqnarray}
\end{widetext}
where $\tau$ is the time associated with the $x$ cordinate. 
Up to terms involving equal time commutators, $[\Theta_{0\nu}(\tau,{\vec x}),\phi(\tau, {\vec y}_i)]$, generated by 
differentiating the time ordering,  $T_{\mu\nu}(x; y_1, y_2)$ satisfies an equation analogous to eq.~(\ref{conscond}).

Aside from mass renormalizations, the correction to $T_{\mu\nu}(x; y_1, y_2)$ to order $g$ is
\begin{widetext}
\begin{eqnarray}\label{correction}
\delta T_{\mu\nu}(x; y_1, y_2)&=&g\int d^5z\delta(z^2+1) \{\tau^{2-D}\partial_\mu D_m(x,z)\partial_\nu D_m(x,z)\nonumber\\&-&\frac{\eta_{\mu\nu}}{2}[\tau^{2-D}\partial^\alpha D_m(x,z)\partial_\alpha D_m(x,z)-\tau^{-D}m^2D_m(x,z)^2]\}D_m(z,y_1)D_m(z,y_2)\}\, .\nonumber\\
\end{eqnarray}
\end{widetext}
For the de Sitter currents to be conserved we require that
\begin{equation}\label{conscond-2}
\Delta(x;y_1,y_2)=\tau\partial^\mu\delta T_{0\mu}(x; y_1, y_2)-\delta T^{\mu}\, _{\mu}(x; y_1, y_2)=0\, .
\end{equation} 
The validity of 
\begin{eqnarray}\label{formalcond}
&\int& d^5z\delta(z^2+1) \{\tau^{2-D}\partial^{\mu}\partial_\mu D(x,z)\partial_\nu D(x,z)\nonumber\\&-&\tau^{-D}m^2\partial_\nu D(x,z)^2]\}D(z,y_1)D(z,y_2)=0\, ,
\end{eqnarray}
which follows from the equations of motion, would insure (\ref{conscond-2})
Although this relation is {\em formally} satisfied it involves products of functions singular, both, at $x=z$ and at $x={\bar z}$;  thus before we conclude anything the integral in (\ref{conscond-2}) must be regulated. As the singularity at $x=z$ is a short distance one, the curvature of the underlying space does not come into play and it is removed by the ususl ultraviolet renormalization. The singularity at $x={\bar z}$ is new and requires its own regularization. 

As the residue of the pole at $z_{12}=-1$ in (\ref{polyakovprop}) does depend on the mass \cite{Erdelyi}, the regularization we use consists of {\em subtracting} from (\ref{correction})  an expression in with all propagators $D_m(x,z)=(1-z_{12}^2)^{-(D-2)/4}{\cal Q}_{-\frac{1}{2}+i\nu(m)}^{(D-2)/2}(z_{12})$ replaced by 
\[
D_M(x,z)=\left[{\cos(i\nu(m)+(D-2)/2)}/{\cos(i\nu(M)+(D-2)/2}\right]
(1-z_{12}^2)^{-(D-2)/4}{\cal Q}_{-\frac{1}{2}+i\nu(M)}^{(D-2)/2}(z_{12})
\]
 (the prefactor involving the cosines makes the residues at $z=-1$ in $D_m$ and $D_M$ equal) and at the end letting $M\rightarrow\infty$. The substituion, $m\rightarrow M$ is performed only in the propagators and not in the $m$ that appears explicitly in (\ref{correction}). The formal manipulations may now be carried out resulting in
\begin{widetext}
\begin{equation}\label{anomaly-1}
\Delta(x;y_1,y_2)=\int d^5z\delta(z^2+1)\tau^{1-D}(M^2-m^2)D_M(x,z)^2D_m(z,y_1)D_m(z,y_2)\, .
\end{equation}
\end{widetext}
The conservation of the de Sitter currents depends on whether $\Delta\rightarrow 0$ as $M\rightarrow\infty$.  To perform the above integration we follow the procedure of \cite{Polyakov:2009nq} which we outline here. Details will be presented elsewhere \cite{MBHR}. We shall show that in the large $M$ limit the integrand will be peaked at $z={\bar x}$ and we can replace the propagators $D_m(z,y_i)$ by $D_m({\bar x}, y_i)$. Due to Lorentz invariance in the imbedding space, the resultant integral does not depend on $x$ and we may set $x=(0,1,z_{i>1}=0)$ and as $D_M(x,z)=D_M(x\cdot z-i\epsilon)$ (the dot product being taken in the imbedding space) (\ref{anomaly-1}) becomes
\begin{widetext}
\begin{equation}\label{anomaly-2}
\Delta(x;,y_1,y_2)\sim M^2\int dz_0dz_1 (z_0^2-z_1^2+1)_+^{(D-3)/2}D_M(z_1-i\epsilon)^2D_m({\bar x},y_1)D_m({\bar x},y_2)\, ;
\end{equation}
\end{widetext}
the $+$ suscript in $(\cdots)_+$ denotes that the integration is to be restricted to the region where the expression inside the parenthesis is positive. The subsequent $z_0$ integration as well as the use of the analyticity of $D_M(z_1)$ in the lower half plane are discussed in \cite{Polyakov:2009nq}. Repeating that procedure results in
\begin{widetext}
\begin{equation}\label{anomaly-4}
\Delta(x;,y_1,y_2)\sim M^2\int_{-1}^1 dz_1\left\{\begin{array}{cc}
                                                                                                      -2i\pi\epsilon(z_1) & D=2\\
                                                                                                       2(1-i\epsilon(z_1))\sqrt{1-z_1^2} & D=3\\
                                                                                                       2i\pi\epsilon(z_1)(z_1^2-1) & D=4
                                                                                               \end{array}\right\}
D_M(z_1-i\epsilon)^2D_m({\bar x},y_1)D_m({\bar x},y_2)\, .
\end{equation}
\end{widetext}

Using the propagator in (\ref{polyakovprop}),  $D_M(z)=(1-z^2)^{-(D-2)/4}{\cal Q}_{-\frac{1}{2}+i\nu(M)}^{(D-2)/2}(z-i\epsilon)$  we are asked to look at the large $M$ limit of $\left[\cos(i\nu(M)+(d-2)/2\right]D_M(z-i\epsilon)$. From \cite{Erdelyi} we find
\begin{eqnarray}\label{Q-limit}
&\cos[i\nu(M)+(D-2)/2]^{-1}D_M(z)\rightarrow \nonumber\\
&\sqrt{\frac{\pi}{2}}e^{-M\pi}M^{(D-3)/2}\left[\frac{z+(z^2-1)^{\frac{1}{2}}]^{i\nu(M)+\frac{1}{2}}}{(z^2-1)^{\frac{1}{4}}}\right]\, ;
\end{eqnarray}
with all $z$'s having a small negative imaginary part.
At $z=-1$ this limit is infinite while for all $z>-1$ it is zero justifying the replacemrnt of $z$ by ${\bar x}$ in the propagators $D_m(z,y_i)$. The integral of the square of {\ref{Q-limit}) multiplied by the dimendion dpendent factors in (\ref{anomaly-4}) behaves as $M^{-2}$ resulting in 
\begin{equation}\label{anaomaly-5}
\Delta(x;,y_1,y_2)\sim D_m({\bar x},y_1)D_m({\bar x},y_2)\, ,
\end{equation}
or, going back to eqs.~(\ref{dSgena}--\ref{dSgend})
\begin{equation}\label{result-1}
\eta^{\mu\nu}\partial_{\mu}S^{(\cdot,i)}_\nu(x)\sim\left(\tau\frac{\partial}{\partial\tau}+x_i\frac{\partial}{\partial x_i}\right) g\phi({\bar x})^2\, .
\end{equation}
This can be generalized to any interaction of scalar fields as 
\begin{equation}\label{result-2}
\eta^{\mu\nu}\partial_{\mu}S^{(\cdot,\cdot)}_\nu(x)\sim\left(\tau\frac{\partial}{\partial\tau}+x_i\frac{\partial}{\partial x_i}\right) \frac{\partial^2}{\partial\phi^2}V(\phi({\bar x}))\, .
\end{equation}
As the propagator for the Euclidian vacuum has no antipodal singularity, these anomalies do not apply for that case. 

Several questions remain unanswered. The technical ones are: (i) Are these results dependent on regularization schemes?, (ii) How do higher order corrections affect these results?, (iii) Do interacting fermions induce a similar anomaly? The results in \cite{Bander:2010pn} would indicate that the answer is no.
A more fundamental question is: is the propagator in \cite{Polyakov:2007mm} the one to use in perturbative calculations on this positive curvature space or does the result presented here serve as annother nail in the coffin of the $\alpha$-vacua \cite{Einhorn-Larsen}\cite{Banks:2002nv}?

I wish to thank Dr.~ E.~Rabinovici, Dr.~A.~Rajaraman and and Dr.~ A.~Schwimmer for discussions and suggestions. 


\end{document}